\documentstyle[aps,prl,multicol,epsf]{revtex}

\def\bra#1{\langle #1 |}
\def\ket#1{| #1\rangle}
\def\braket#1#2{\langle \, #1 \, | \, #2 \, \rangle}

\def\R{\hbox{\rm I \kern-5pt R}}

\def\Tr{{\rm{Tr}}}
\title{Quantum Bit String Commitment}

\author{Adrian Kent} 
\address{ Hewlett-Packard Laboratories, Filton Road,\\
Stoke Gifford, Bristol BS34 8QZ, U.K.\\
{\it and}\\
Centre for Quantum Computation, DAMTP, Centre for Mathematical
Sciences,\\ University of Cambridge, Wilberforce Road, 
Cambridge CB3 0WA, U.K.${}^1$
}
\date{April 2003 (revised)} 

\begin{document} 
\maketitle
\begin{abstract}
A bit string commitment protocol securely commits $N$ classical bits in such
a way that the recipient can extract only $M<N$ bits of information 
about the string.    
Classical reasoning might suggest that bit string commitment implies 
bit commitment and hence, given the Mayers-Lo-Chau theorem, 
that non-relativistic quantum bit string commitment is impossible.  
Not so: there exist non-relativistic quantum bit string 
commitment protocols, with  
security parameters $\epsilon$ and $M$, that allow $A$ to 
commit $N = N(M, \epsilon)$ bits 
to $B$ so that $A$'s probability of successfully 
cheating when revealing any bit 
and $B$'s probability of extracting more than $N'=N-M$ bits of information 
about the $N$ bit string before revelation are both less than $\epsilon$.   
With a slightly weakened but still restrictive definition of 
security against $A$, $N$ can be taken to be $O( \exp ( C N' ) )$ 
for a positive constant $C$.   
I briefly discuss possible applications.  
\end{abstract}
\vskip 5pt
${}^1$ Present and permanent address.

\begin{multicols}{2}

\section{Introduction} 

As is by now well known, quantum information can guarantee 
classically unattainable security in a variety of important 
cryptographic tasks.  Some no-go results have also been
obtained, showing that quantum cryptography cannot guarantee
perfect security for every task.  We do not presently have a good
characterisation of the tasks for which perfectly secure quantum 
protocols exist.  In fact, we are not 
yet even able to characterise the range of cryptographic 
tasks for which perfectly secure quantum protocols {\it might 
possibly} exist.  The main reason is that quantum cryptography involves
more than devising quantum protocols for tasks known to be useful
in classical cryptography.  The properties of quantum information 
allow one to devise new and cryptographically useful tasks, which
have no classical counterpart.  Moreover, reductions and relations
between classical cryptographic tasks need not necessarily apply 
to their quantum equivalents.  This means that there is
a wider range of tasks to consider, and that no-go theorems may 
not necessarily be quite as powerful as classical reasoning would suggest. 

These remarks apply particularly to bit commitment, an important
cryptographic protocol whose potential for physically secure
implementation has been extensively 
investigated\cite{BBef,BCJL,brassardcrepeau,lochauprl,mayersprl,mayerstrouble,lochau,mayersone,bcms,kentrel,kentrelfinite,kentbccc}.
It is known that unconditionally secure quantum bit commitment is
impossible for non-relativistic 
protocols\cite{lochauprl,mayersprl,mayerstrouble,lochau,mayersone}:
that is, protocols in which the two parties are restricted to 
single pointlike sites, or more generally, in which the 
signalling constraints of special relativity are ignored.   
On the other hand, unconditionally secure bit commitment is 
thought to be possible between parties controlling 
appropriately separated pairs of sites, when the impossibility of 
superluminal signalling is taken into account.\cite{kentrel,kentrelfinite}

While sustaining a bit commitment indefinitely via relativistic
protocols is practical with current technology\cite{kentrelfinite}, 
the constraints it imposes are not always desirable.  Both parties have 
to maintain separated secure locations, and communications
have to continue throughout the duration of the commitment.  
A further motivation for continued study of non-relativistic protocols
is that it is theoretically interesting to characterise which secure quantum
protocols can be implemented without relying on relativity.  
With these motivations in mind, we restrict attention to 
non-relativistic protocols in the rest of this paper. 
Rather than insert the word ``non-relativistic'' throughout, 
we generally take the restriction as understood below.  

Some variants of bit commitment, for which non-relativistic
protocols are not known to be impossible, have previously been  
studied.\cite{hk,atvy} 
This paper considers a different type of generalisation, bit string 
commitment, in which one party commits many bits to another in a 
single protocol.  Two non-relativistic bit string commitment 
protocols, which offer classically
unattainable levels of security against cheating, are described.  

\section{Bit string commitment}  

Consider the following classical cryptographic problem. 
Two mistrustful parties, A and B, need a protocol which
will (i) allow A to commit a string $a_1 a_2 \ldots a_n$ of 
bits to B, and then, (ii) at any later time of her choice, 
reveal the committed bits.   The protocol should prevent $A$ 
from cheating, in the sense that she should have negligible
chance of unveiling bits $a'_i$ different from the 
$a_i$ without B being able to detect the attempted detection.
In other words, A should be genuinely committed after the
first stage.   The protocol should also prevent $B$ from
being able to completely determine the bit string.
More precisely, it must guarantee that, before revelation,
$B$ has little or no chance of obtaining more than $m$ bits of information 
about the committed string, for some fixed integer $m < n$.   

This {\it $(m,n)$ bit string commitment} problem
is a generalisation of the standard bit commitment problem, 
for which $n=1$ and $m=0$.  Clearly, a protocol for bit commitment
would solve this generalised problem, since the protocol could 
be repeated $n$ times to commit each of the $a_i$, and $B$ would
be able to obtain no information about the committed string.   
Conversely, classical reasoning implies that a protocol for the 
generalised problem, for any integers $m$ and $n$ with $m<n$, could be 
used as a protocol for standard bit commitment.  For $A$ and $B$ could
use any coding of a single bit $a = f ( a_1 , \ldots , a_n )$ 
in terms the $n$ bit string such that
none of the $m$ bits available to $B$ is correlated with $a$, 
and then use the protocol to commit $A$ to $a$.   

Classically, then, $(m,n)$ bit string commitment is essentially
equivalent to bit commitment.  At first sight, allowing $A$ and
$B$ to use quantum information may seem to make no difference.
But there are subtleties.  One is that extracting information from a quantum 
state can generally be done in many different ways.  Each of 
these generally disturbs the quantum state, so that different 
ways of information extraction are generally incompatible: after
method one has been applied, method two may no longer give as 
much (or any) information.  This leaves open the possibility of 
bit string commitment protocols in which $B$ can obtain some $m$
bits of information about the committed $n$ bit string in many different
ways, without $A$ necessarily knowing
which $m$ bits of information are obtained. 
A second subtlety is that if $A$ commits a mixed state, a 
protocol can leave her almost 
perfectly committed to each bit in a string, in the sense that
she is essentially unable to vary the probabilities of revealing
$0$ or $1$ for any given bit, while leaving the actual bit values
undetermined until a measurement at the revelation stage. 
For a long enough string, this might be doable
in such a way as to leave $A$ almost 
completely {\it uncommitted} to the value of some
joint functions of the string bits.  

Any attempt to use a secure quantum bit string protocol to commit a
single bit by redundant coding could thus fail: it could be that, for
any given coding, either $A$ or $B$ can cheat.  In other words, there
is no obvious equivalence between quantum $(m,n)$ bit string
commitment and quantum bit commitment.  The impossibility of
unconditionally secure quantum bit commitment does not necessarily
imply that unconditionally secure quantum bit string commitment, with
an analogous definition of security, is impossible.  
We now show it can be achieved.

\section{Protocol 1} 

Define qubit states $\psi_0 = \ket{0}$ and $\psi_1 = \sin \theta \ket{0} +
\cos \theta \ket{1}$.  We take $\theta > 0 $ to be small;
$\theta$ and $r=n-m$ are security parameters for the
protocol.   

{\bf Commitment:} \qquad 
To commit a string $a_1 \ldots a_n$ of bits to $B$, $A$ sends the 
qubits $\psi_{a_1} \, , \ldots \, , \, \psi_{a_n}$, sequentially.
\vskip 10pt
{\bf Unveiling:}\qquad 
To unveil, $A$ simply declares the values of the string bits, and
hence the qubits sent.  Assuming that $B$ has not disturbed the
qubits, he can test the bit values $a'_i$ claimed
by $A$ at unveiling by measuring the projection onto $\psi_{a'_i}$ 
on qubit $i$, for each $i$.  If he obtains eigenvalue $1$ in each
case, he accepts the unveiling as an honest revelation of a genuine
commitment; otherwise he concludes $A$ cheated.  
(As usual, we assume noiseless channels.) 
\vskip 10pt
{\bf Security against A:} \qquad  Whatever strategy $A$ follows, 
once she transmits the qubits to $B$, their respective density
matrices $\rho_i$ are fixed.  Let $p^j_i = \bra{\psi_j} \rho_i \ket{\psi_j }$
be the probability of $B$ accepting a revelation of $j$ for the $i$-th
bit.  We have 
\begin{equation}
p^0_i + p^1_i \leq \cos^2 ( (\pi - 2 \theta ) / 4 )  + 
               \sin^2 ( ( \pi  + 2 \theta ) / 4 )  \, , 
\end{equation}
which is $\leq 1 + \theta $ for small $\theta$.  
This is the standard definition of security against $A$ for an individual bit
commitment, with security parameter $\theta$.  
In other words, $A$'s scope for cheating on any bit of the string
is limited to slightly increasing the probability of revealing 
a $0$ or $1$, by an amount $\leq \theta$, which can be made arbitrarily
small by choosing the security parameters appropriately. 
$A$ is committed to each individual bit, in this standard sense,
although of course the protocol does not prevent her committing
quantum superpositions of bits or bit strings.  
\vskip 10pt
{\bf Security against B:} \qquad  
We assume that prior to commitment $B$ has no information about 
the bit string: to $B$, all string values are equiprobable.   
He thus has to obtain information about
a density matrix of the form 
\begin{equation}
\rho =  2^{-n} \sum_{a_1 \ldots a_n} \ket{ \psi_{a_1} \ldots \psi_{a_n} }
\bra{ \psi_{a_1} \ldots \psi_{a_n} } \, .
\end{equation}
Holevo's theorem\cite{holevo} 
tells us that the accessible information available to
$B$ by any measurement on $\rho$ is bounded by the entropy 
\begin{equation}
 S ( \rho )= (H ( \frac{1 + \sin \theta}{2} ))^n 
\end{equation}
For any fixed $\theta > 0 $, we have $S(\rho ) < n $.  For any
fixed $r$, by taking $n$ sufficiently large, we can ensure 
$ n - S( \rho ) > r$.   So we can ensure that,
however $B$ proceeds, on average at least $r$ bits of information
about the string will remain inaccessible to him.  

By choosing $n$
suitably large, we can also ensure that the probability of $B$
obtaining more than $n-r$ bits of information about the string is 
smaller than any given $\delta >0$.  A simple bound follows from
considering the probability of $B$
identifying all $n$ bits.  As each bit is equiprobably
$0$ or $1$, $B$'s probability of identifying it is
no more than $H ( \frac{1 + \sin \theta}{2} )$; his probability of identifying
all $n$ is no more than $ ( H ( \frac{1 + \sin \theta}{2} ) )^n $. 
If he obtains more than $n-r$ bits of information about the string, his
probability of identifying all $n$ bits is greater than $2^{-r}$.
Hence $ \delta \leq 2^r ( H ( \frac{1 + \sin \theta}{2} ) )^n $. 

\section{Protocol 2} 

Protocol 1 ensures bit-wise security against $A$, but uses 
a rather inefficient bit string coding which allows $B$ to obtain 
almost all of the bit string before revelation.     
For large $n$, more efficient codings allow the security against $B$ 
to be greatly enhanced, though with a weakened notion of security
against $A$.   

We again take the security parameter $\theta > 0$ to be small
and write $\epsilon = \sin \theta$.
For any $\theta > 0$ and sufficiently large $n$,
explicit constructions are known for sets $v_1 , \ldots, v_{f(n)}$ of
vectors in $H^n$ such that $ | \braket{ v_i}{ v_j } | < \sin \theta $
for all $i \neq j$, with the property that $f(n) = O ( \exp ( C n )
)$, where $C$ is a positive constant that depends on
$\theta$.\cite{conwaysloane,justesen} (The use of these constructions
for efficient quantum coding of classical information has previously
been noted by Buhrman et al.\cite{bcww}, who describe efficient
quantum fingerprinting schemes which reduce communication complexity
in the simultaneous message passing model.)  A string of $O( C n )$
bits can thus be encoded by vectors in $H^n$, such that the overlap
between the code vectors for two distinct strings is always less than
$\sin \theta$, suggesting the following bit string commitment
protocol.

{\bf Commitment:} \qquad 
Let $N$ be the number of bits that can be encoded in $H^n$ by the
above construction.  To commit a 
string $a_1 \ldots a_N$ of bits to $B$, $A$ sends the 
state $v_{a_1 \ldots a_N}$, treating the index as a binary number.
\vskip 10pt
{\bf Unveiling:}\qquad 
To unveil, $A$ declares the values of the string bits, and
hence the state sent.  Assuming $B$ has not disturbed the
qubits, he can test $A$'s claim by measuring the projection
onto $v_{a_1 \ldots a_N}$.  If he obtains eigenvalue $1$, he accepts 
the unveiling as an honest revelation of a genuine
commitment; otherwise he concludes $A$ cheated.  
\vskip 10pt
{\bf Security against A:} \qquad  As before, 
once $A$ transmits a quantum state to $B$, its density
matrix $\rho$ is fixed.  
Consider some set $i_1 , \ldots , i_r$ of bit strings which 
$A$ might wish to maintain the option of revealing
after commitment.  Let $P_i$ be the projection onto $v_i$, let $p_i = 
\Tr (\rho P_i )$ be the probability of $A$ successfully revealing 
string $i$, 
and write 
\begin{equation} Q = P_{i_1} + \ldots + P_{i_r} \, . \end{equation}

We want to bound $\Tr ( \rho Q) $ for any density matrix $\rho$. 
This can be done by first maximising
$\langle Q \rangle_w = \frac{  \bra{w} Q \ket{w} }{ \braket{w}{w}}$ for any
vector $\ket{w}$.   Writing $\ket{w} = \sum_j w_j \ket{v_{i_j}} +
\ket{v^{\perp}}$, where $\braket{ v^{\perp}}{v_{i_j}} = 0 $ for 
$j$ from $1$ to $r$, clearly $\ket{v^{\perp}} = 0$ maximises 
$\langle Q \rangle_w$.   So without loss of generality we can
write $\ket{w} = \sum_j w_j \ket{v_{i_j}}$ with $\sum_j | w_j |^2 =
1$.

Now 
$$\langle Q \rangle_w = \frac{ 1 + 2 S_2 + S_3} { 1 + S_2}
\, , $$
where 
$$ S_2 = \sum_{ij} \bar{w_i } w_j ( 1 -
  \delta_{ij} ) \braket{v_i }{v_j } $$
and
$$ S_3 = 
 \sum_{ijk} \bar{w_i } w_k ( 1 -   \delta_{ij} ) ( 1 - \delta_{jk} )
\braket{v_i }{v_j } \braket{v_j}{v_k } \, . $$
The Cauchy-Schwarz inequality gives us that 
$$
S_2 \leq \epsilon (r-1) \, \qquad S_3 \leq \epsilon^2 (r-1)^2
\, . $$
Both bounds are simultaneously attainable, by setting $w_j =
\frac{1}{\sqrt{r}}$ for all $j$ and $\braket{v_{i_j}}{v_{i_k}} =
\epsilon$ for all $j,k$.  
Also, it is easy to see that, provided $(r-1) \epsilon < 1$ (which we
assume), the maximum of $\langle Q \rangle_w $ is attained when
$S_2$ and $S_3$ are simultaneously maximised.  (Geometrically, the
largest possible eigenvalue of $Q$ arises when the $v_{i_j}$ bunch
as closely as possible, and then the corresponding eigenvector is the
sum of the $v_{i_j}$.) 
We thus have that 
$$\langle Q \rangle_w \leq 1 + (r-1) \epsilon \, .$$

More generally, since any state $\rho$ can be written as
a mixture of pure states, we have for all states 
\begin{equation} 
\Tr ( \rho Q ) \leq 1 + (r-1) \epsilon \, .
\end{equation}
In other words, 
\begin{equation}
p_{i_1} + \ldots + p_{i_r} \leq 1 + (r-1) \epsilon \, , 
\end{equation}
and for any fixed $r$, this can be made as close to $1$ as desired 
by choosing $\theta$ suitably small.  

So, if $A$ is determined to reveal a bit string from
some finite set of size $r$, her scope for cheating is limited to increasing
the probability of revealing any given element of the set
by a fixed amount.  For any fixed $r$, that amount can be made 
arbitrarily small by choosing the security parameters appropriately. 
If $B$'s concern is to prevent cheating of this type, for some
predetermined $r$, the protocol can guarantee him security. 
\vskip 10pt
{\bf Security against B:} \qquad  Holevo's theorem implies
that the information about the $N \approx C n$ bit string accessible to
$B$ is at most $\log n$ bits.   

\section{Discussion} 

The bit string commitment protocols above use the properties
of quantum information to guarantee strong
levels of security to both the committer and receiver.
They highlight another cryptographic application of quantum
information: no (non-relativistic) classical protocol can guarantee 
such security.  They also highlight the fact that quantum cryptography
can introduce distinctions between tasks which are classically 
equivalent, such as bit commitment and bit string commitment.  

As a metaphor for the cryptographic uses of bit string commitment
--- in particular, of the second type of protocol --- 
consider a situation in which $A$ knows the combination to a lock,
wants to be able to prove to $B$ in future that she knows it now, 
but does not want to give $B$ the ability to open the lock now.  
If she sends a bit string commitment of the combination now, she can
prove her present knowledge later by opening the commitment.  
However, $B$, who can only get partial information about the
committed string, will not be able to deduce the combination 
from it.  If the combination is sufficiently long, the
security parameters for the bit string commitment are appropriately 
chosen, and $A$ knows how fast $B$ can try possible combinations,
she can ensure that $B$ remains sufficiently ignorant
about the combination to be almost certainly unable to break the lock 
during some fixed interval of her choice.  

As another illustration, suppose that $A$ has just 
obtained a very high resolution image
of something of interest to, but kept secret from, $B$.    She 
may wish to be able
to prove to $B$ later that she had the image today --- so that he 
will take her seriously enough to purchase her services in future --- 
without revealing too much detailed information to $B$ for free.  
A quantum bit string commitment protocol with suitable parameters
could meet this need.  

One might think that both these applications could also be implemented
securely classically, simply by allowing $B$ to choose a random
subset of the combination or image and asking $A$ to provide the
data corresponding to that subset.  However, this would persuade
$B$ only that $A$ is able to compute or obtain a dataset of the
size of the subset.  She might be able to do this with a device
that extracts the combination digit by digit, or with an imaging
device of restricted field, without actually being able to 
obtain all the data at the time she claims to have it.  

More generally, bit string commitment allows a sort 
of partial knowledge proof, in which $A$ can establish to 
$B$ her possession of some information --- the factorisation of
a number, the proof of a theorem, $\ldots$ --- while restricting 
the amount of information $B$ can obtain.    It also illustrates
the general possibility in quantum cryptography of iterating a 
protocol a number of times with a partial security guarantee
that allows the parties to be certain that many or most of the bits 
involved are appropriately controlled. Practical 
cryptographic applications that require bit commitment almost
always involve strings of bits, and perfect security of
the entire string may often not be essential.
Moreover, quantum bit string commitment can be used on top
of classical bit commitment schemes, offering an
extra layer of classically unobtainable 
security with a partial but unconditional
security guarantee.  It thus seems likely to be rather useful. 

\section{Acknowledgements} 

This work was partly supported by the European collaborations EQUIP
and PROSECCO.  I thank Serge Massar and Alain Tapp for helpful discussions.

\end{multicols} 
\end{document}